\documentclass[preprint,12pt]{elsarticle}

\usepackage{graphicx,stfloats}
\usepackage{color}
\usepackage{lineno}
\usepackage{fixmath}
\usepackage{amssymb}

\usepackage{amsmath}
\DeclareMathOperator*{\argmax}{arg\,max}
\DeclareMathOperator*{\argmin}{arg\,min}

\biboptions{sort&compress}

\journal{Proceedings of the Combustion Institute}

\begin{document}

\begin{frontmatter}

\title{Information-Driven Design for Shock Tube / Laser Absorption Studies of Fundamental Rate Constants in Combustion, with Application to Methanol Pyrolysis}

\author{Shengkai Wang\corref{cor1}}
\ead{sk.wang@stanford.edu}

\author{Yiming Ding}
\author{Ronald K. Hanson}

\address{High-Temperature Gasdynamics Laboratory, Stanford University, CA, 94305, USA}
\cortext[cor1]{Corresponding author:}

\begin{abstract}
Shock tube experiments, paired with precision laser diagnostics, are ideal venues to provide kinetics data critically needed for the development, validation and optimization of modern combustion kinetics models. However, to design sensitive, accurate, feasible and information-rich experiments that may yield such data often requires sophisticated planning. This study presents a mathematical framework and quantitative approach to guide such experimental design, namely a method to pin-point the optimal conditions for specific experimentation under realistic constraints of the shock tubes and diagnostic tools involved. For demonstration purpose, the current work is focused on a key type of shock tube kinetic experiments -- direct determination of fundamental reaction rate constants. Specifically, this study utilizes a Bayesian approach to maximize the prior-posterior gain in Shannon information of the rate constants to be inferred from the intended experiment. Example application of this method to the experimental determination of the CH$_3$OH + H (k$_1$) and CH$_2$O + H (k$_2$) rate constants is demonstrated in shock tube/laser absorption studies of the CH$_3$OH pyrolysis system, yielding new recommended rate constant expressions (over 1287 K - 1537 K) as: k$_1$ = $ 2.50 \times 10^6 (T/K)^{2.35} exp(-2975 K /T) \, cm^3mol^{-1}s^{-1} \pm 11.4\%$ and k$_2$ = $7.06 \times 10^7 (T/K)^{1.9} exp(-1380 K/T) \, cm^3mol^{-1}s^{-1} \pm 9.7 \%$. Potential extension to other types of kinetic studies, e.g. prediction of combustion benchmarks such as ignition delay times and species yields, and global uncertainty minimization of generic reaction models, are also briefly discussed.
\end{abstract}


\end{frontmatter}

\section{Introduction}

There has been a strong and continuous need for accurate and relevant experimental data to aid development, validation and optimization of gas-phase combustion kinetics models, particularly for use in practical and realistic combustion applications. Recent advancements in shock tube methods and supporting diagnostics (e.g. high-speed mass spectrometry \cite{tranter2007shock,durrstein2011shock,sela2016single}, photo ionization mass spectrometry \cite{lynch2015probing}, multi-species laser absorption \cite{davidson2011multi}, high-speed imaging \cite{tulgestke2018high}, ultra-sensitive laser diagnostics using cavity-enhanced spectroscopy \cite{wang2017time,alquaity2015sensitive} and frequency-modulation spectroscopy \cite{wang2018ultra}, and the combination of laser absorption with gas chromatography \cite{ferris2018combined}) have opened pathways to new regimes of combustion kinetic studies previously not accessible, thereby enabling researchers to revisit some long-standing challenges (e.g. low-temperature chemistry), as well as to investigate newly emerging research topics in combustion. It is worth mentioning, however, that the availability of these new methodologies alone does not automatically guarantee the success of such research efforts, which may require sophisticated planning and unique insight about the experimental design. A prerequisite for successful and informative shock tube kinetic experiments is to identify the most relevant properties to be measured and to select sensitive conditions and diagnostics for performing these measurements while minimizing experimental uncertainties.

Of particular interest to the current study is a key type of shock tube kinetic experiments, i.e. direct determination of fundamental reaction rate constants. Conventional wisdom, guided by either global or local sensitivity analysis \cite{miller1990chemical,turanyi1990sensitivity,tomlin2013role}, often suggest such experiments should be performed at "kinetically clean" or "chemically isolated" conditions where a small number (usually one or two) of reactions are dominant, for example, reaction systems with heavily-diluted reactants where the influences of secondary reactions can be suppressed \cite{wang2016shock}. However, such conditions are generally hard to meet due to various experimental constraints (e.g. detection limits of diagnostics), and for most situations, certain trade-offs between the kinetic sensitivity and signal strength of an experiment are needed. Historically, kinetics researchers have relied largely on experience or a trial-and-error approach to find the optimal balance between the two, due largely to the lack of quantitative theory and a systematic framework that can rigorously determine the conditions corresponding to the optimal experimental design. This background strongly motivated the current study. Specifically, the current work introduces a Shannon information-based Bayesian approach to identify the optimal conditions for specific experimentation under realistic constraints of facility (e.g. non-ideal effects in shock-heated gases) and diagnostic tools (e.g. resolution limits of laser absorption sensors) involved. 

It is worthwhile to note the difference between the current study and previous efforts along a related yet different line of research that also aimed to maximize the knowledge gain from experiments. Those efforts focused mostly on extraction of useful information from given experimental data, i.e. model optimization and uncertainty quantification (UQ). The pioneering work of adapting UQ to systematic optimization of combustion models can be traced back to the 1980s \cite{miller1983sensitivity}; over the last three decades, significant success has been achieved in tailoring UQ to combustion chemistry, and now it has become a standard tool for quantitative evaluation and optimization of complex reaction models. For a historical prospective of the research progress and representative work along this path, the reader is referred to a comprehensive review paper by Wang and Sheen \cite{wang2015combustion}. The scope of the current work, on the other hand, is focused on the "inverse problem" of UQ, i.e. for a given trial model and prior knowledge about the uncertainty of the model (expressed in term of the joint probability distribution of individual rate constants), to design the optimal experiment within physical and facility constraints that would provide the most information for selected rate constants. In the authors' view, properly addressing this "inverse problem" is equally important as UQ, because high-quality experiments are the very foundation that model validation, optimization and uncertainty quantification is based on. Besides, the systematic experimental design approach proposed in the current work, when harnessed together with UQ, can be instrumental in the development of predictive modeling.

\section{Method}
\subsection{Quantifying the Worth of an Experiment}
Mathematically, shock tube measurement of fundamental rate constants (one or several, here denoted collectively as \{$k_i$\}) is essentially an inference problem -- one performs a set of shock tube kinetic experiments in which the direct observable is a strong function of the target rates, and infers the most probable rate value that fits the experimental data (denoted as \{$d_i$\}). Typical data types of such experiments include global data such as ignition delay times \cite{shao2019shock} and species yields at a given time (or peak values) \cite{sarathy2012comprehensive,zhang2018acetaldehyde}, and time-resolved data such as species concentration and temperature profiles/histories \cite{hong2011new,hanson2013constrained}. For demonstration purpose, the current work focuses on laser absorption measurement of species time-histories, as it contains dense information that can typically provide the strongest constraints to chemical kinetic models. Nonetheless, the generic method developed in the current work can be easily extended to other data types, or to the combination of several types of data as well.

\begin{figure}
    \centering
    \includegraphics[width=0.8\linewidth]{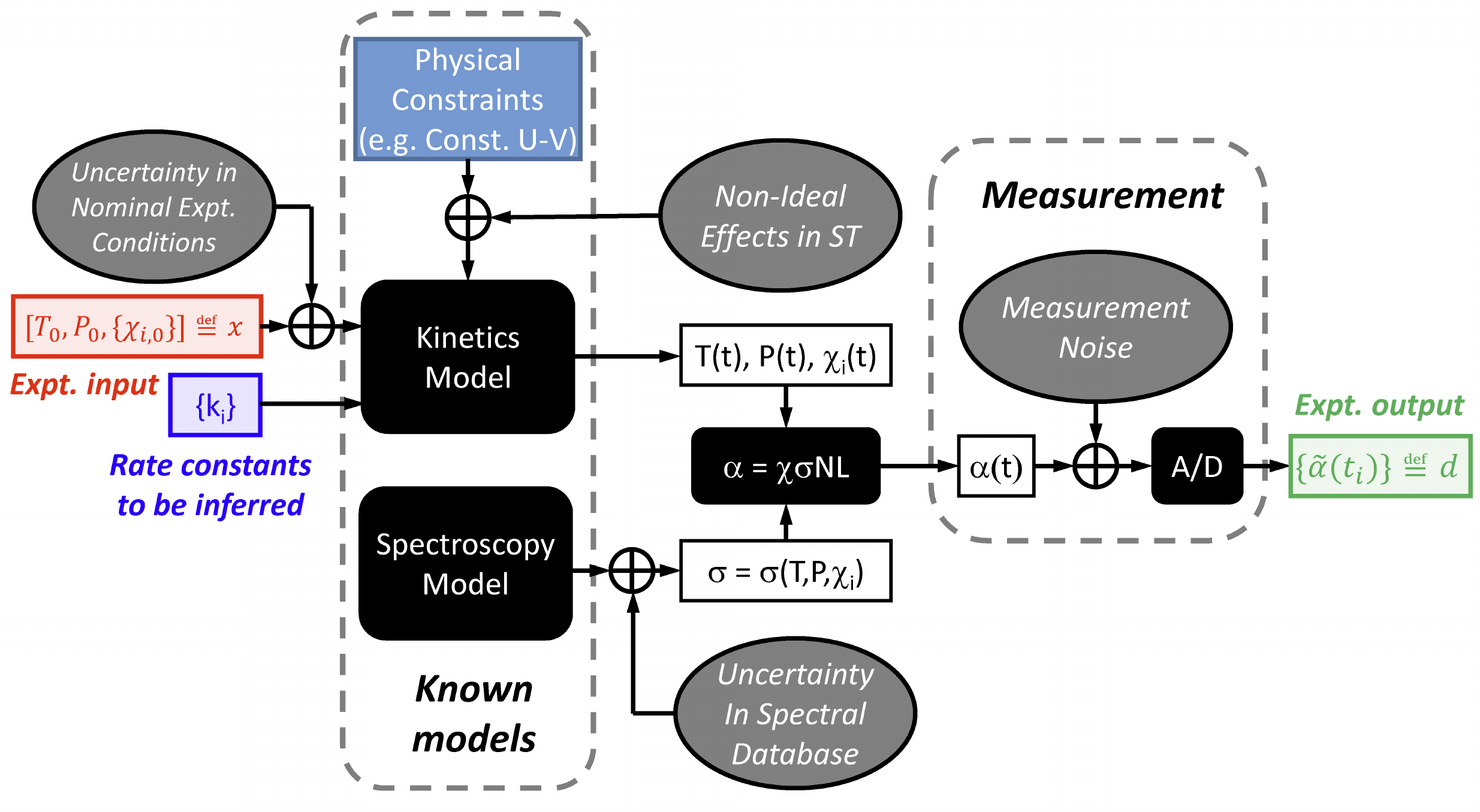}
    \caption{Information flow and uncertainty sources in typical shock tube and laser absorption measurements of fundamental reaction rate constants}
    \label{fig:F1}
\end{figure}

A schematic diagram of the information flow in a typical shock tube / laser absorption experiment for determining fundamental rate constants is shown in Fig. \ref{fig:F1}. In such an experiment the physical and chemical processes being studied are assumed to be governed by two models -- a kinetics model and a spectroscopy model. The kinetics model predicts the time-histories of gas temperature ($T$), pressure ($P$) and composition ($\{\chi_i\}$) based on the initial experimental condition ($T_0$, $P_0$ and $\{\chi_{i,0}\}$, here defined collectively as the experiment input variable, $\mathbold{x}$), a set of rate constants ($\{k_i\}$) and given physical constraints of the reaction system (e.g. constant internal energy and volume for typical shock tube experiments). The spectroscopy model predicts the absorbance time-history $\alpha(t)$ using the outputs of the kinetics model based on the Beer-Lambert law. For fractional transmission of monochromatic light of wavelength $\lambda$, $(I/I_0)_\lambda  = exp(-\alpha)$, where $\alpha = \chi \sigma n L$ is the absorbance, $\sigma$ is the absorption cross-section of the species being probed, $\chi$ is its concentration, $n$ is the total gas number density, and $L$ is the optical path-length that usually equals the inner diameter of the shock tube. The physical measurement samples $\alpha(t)$ at finite sampling rate and yields a discrete time sequence through digital data acquisition. The result is defined as the output variable of the experiment, $\mathbold{d}$. The objective of the experiment is to infer $\{k_i\}$ from a set of $\mathbold{d}(\mathbold{x})$, which is in the opposite direction of the actual flow of information.
Conventionally, the success of such experiment is measured by the uncertainty in the inferred rate constants (or the joint probability distribution of rate constants \cite{wang2017shock}), which can only be evaluated after the experiments. To predict the benefit of an experiment prior to its execution, the current study adopts a Bayesian probabilistic approach. Since the probability distribution of rate constant uncertainty is commonly treated as being lognormal \cite{tao2019joint}, the current work uses the notation of normalized rate constants, $\kappa_i = ln(k_i/k_{i,0})$, where $k_{i,0}$ is the reference value for $k_i$. Here we further define $\mathbold{K}$ and $\mathbold{D}$ as the random variables (generally muti-dimensional and expressed in vector form) representing the distribution of $\mathbold{\kappa}$ and $\mathbold{d}$, $prior(\mathbold{K})$ and $prob(\mathbold{K}|\mathbold{D})$ as the prior and posterior probability density function (PDF) of $\mathbold{K}$, and $prob(\mathbold{D}|\mathbold{K})$ as the conditional probability density of $\mathbold{D}$ for a given $\mathbold{K}$, respectively. The current work also assumes a Gaussian prior PDF, i.e. $prior(\mathbold{K}) \sim \mathcal{N}(\textbf{0},\mathbold{\Sigma_\kappa})$, with $\mathbold{\Sigma_\kappa}$ being the co-variance matrix of normalized rate constants. Based on Bayes' Theorem, 
\begin{equation}
\label{eq1}
prob(\mathbold{K}|\mathbold{D}) = \frac{prob(\mathbold{D}|\mathbold{K})prior(\mathbold{K})}{\int prob(\mathbold{D}|\mathbold{K})prior(\mathbold{K})d\mathbold{K}}
\end{equation}
A good experiment would be one that yields a posterior PDF more localized than the prior. To quantify the prior-posterior knowledge gain, the current work adopts the mathematical principle of information theory. Specifically, the knowledge gain in the PDF of $\mathbold{K}$ from a given data set $\mathbold{D}$ is measured by the Shannon information (in the unit of bit) \cite{shannon1948mathematical}:
\begin{equation}
\label{eq2}
i(\mathbold{D}) = \int{prob(\mathbold{K}|\mathbold{D})log_2\Big[\frac{prob(\mathbold{K}|\mathbold{D})}{prior(\mathbold{K})}\Big]d\mathbold{K}}
\end{equation}
The overall benefit or worth of an experiment is measured by the mathematical expectation of the Shannon information, i.e.
\begin{equation}
\label{eq3}
I(\mathbold{x}) = E[i(\mathbold{D})]= \int{i(\mathbold{D})prob(\mathbold{D})d\mathbold{D}}
\end{equation}

Accurate evaluation of $I(\mathbold{x})$ requires comprehensive description of $D(x;\kappa)$ and detailed analysis of experimental uncertainties. Typical sources of uncertainties in shock tube / laser absorption measurements of rate constants can be grouped into three primary categories, namely the random noise in the absorption signal, systematic uncertainties in the experiments and influences of interfering reactions. The first type of uncertainty usually results from electrical noise of the detectors and quantization noise of the digital data acquisition system, and can be modeled as a random vector following Gaussian distribution, i.e. $\mathbold{\epsilon_y} \sim \mathcal{N}(\textbf{0}, \mathbold{\Sigma_y})$, where $\mathbold{\Sigma_y}$ is the co-variance matrix of noise at individual sample points. Due to the finite physical bandwidth of the detection and data acquisition system, this noise is usually temporally correlated. The current study assumes $\mathbold{\Sigma_y} = \sigma^2_y/\eta \cdot \mathbold{I_{N \times N}}$, in which $\sigma_y$ is the $1\sigma$ noise amplitude, $\mathbold{I_{N \times N}}$ is the identity matrix, $N$ is the total number of data points in the measured absorbance time-history, and $\eta \in [1/N, 1]$ is a parameter reflecting the temporal correlation of noise. Essentially, $\eta N$ represents the effective number of independent data points. In the current work, the most conservative assumption of $\eta N = 1$ is adopted to avoid overfitting, as well as to compensate for the potential presence of low-frequency noise and uncorrected systematic drifts (e.g. non-ideal pressure raise behind reflected shock wave) and to ensure robustness of the rate constant determination.

Systematic uncertainties are often more complex as their influences on the measurement results are generally nonlinear. This category of uncertainties includes: (1) non-ideal gasdynamic effects (e.g. shock wave - boundary layer interactions) that may lead to small variations of temperature and pressure behind reflected shock waves; (2) uncertainty in the definition of time zero (often blurred by the schlieren spikes upon the arrival of shock waves); (3) uncertainty in the diagnostics systems (uncertainties in the spectral database, interference absorption, small drifts in the signal baseline, etc.); (4) uncertainties in the test mixture composition; (5) residual impurity in shock tube and mixing manifold, etc. The current work treats the systematic uncertainties as latent random variables (denoted collectively as $\mathbold{\xi}$) whose exact value are unknown throughout an experiment. Their PDF, however, is assumed to be a known Gaussian distribution, i.e. $\mathbold{\xi} \sim \mathcal{N}(\textbf{0},\mathbold{\Sigma_\xi})$, and their influences on the measured signal are calculated through linearized uncertainty propagation using the sensitivity matrix, i.e. $\mathbold{S}_\mathbold{\xi} = \partial \mathbold{F_M}/ \partial \mathbold{\xi}$ (evaluated at $\mathbold{\xi} = \textbf{0}$), where $\mathbold{F_M} =\mathbold{F_M}(\mathbold{x};\mathbold{\kappa},\mathbold{\xi}) = \{f_M(t_i,\mathbold{x};\mathbold{\kappa},\mathbold{\xi}) \}$ is the expected absorbance time-history signal predicted by the kinetic and spectroscopy models. The influences of interfering reactions can be treated in the same way as systematic uncertainties. Based on the Gaussian noise assumptions and local linearization around $\mathbold{\xi} = \textbf{0}$, it can be shown that the measured absorbance time-history also follows a Gaussian distribution:
\begin{equation}
\label{eq4}
\mathbold{D}(\mathbold{x};\mathbold{\kappa})= \mathbold{F_M}(\mathbold{x};\mathbold{\kappa},\mathbold{\xi})+\mathbold{\epsilon_y} \sim \mathcal{N}(\mathbold{F_M}(\mathbold{x};\mathbold{\kappa},\textbf{0}),\mathbold{S_\xi\Sigma_{\xi}S}^T_{\mathbold{\xi}}+\mathbold{\Sigma_{y}})
\end{equation}
Based on Eqs. \ref{eq1} - \ref{eq4}, it can be further shown that the expected information gain can be calculated from the determinant of two matrices:
\begin{equation}
\label{eq5}
I(\mathbold{x})= \frac{1}{2} log_2 \Big[\frac{det(\mathbold{S_{\kappa}\Sigma_{\kappa}S}^T_{\mathbold{\kappa}}+\mathbold{S_\xi\Sigma_{\xi}S}^T_{\mathbold{\xi}}+\mathbold{\Sigma_{y}})}{det(\mathbold{S_\xi\Sigma_{\xi}S}^T_{\mathbold{\xi}}+\mathbold{\Sigma_{y}})} \Big] \equiv \frac{1}{2} log_2 \Big[ \frac{det[\mathbold{M_2(x)}]} {det[\mathbold{M_1(x)}]} \Big]
\end{equation}
where $\mathbold{S_\kappa} = \partial \mathbold{F_M}/ \partial \mathbold{\kappa}$ (evaluated at $\mathbold{\kappa} = \textbf{0}$) is the sensitivity matrix of the expected signal to the target (normalized) rate constants. The physical meaning of $\mathbold{M_1(x)}$ can be viewed as the total "momentum" (product of sensitivity and uncertainty factor) of various uncertainty sources, whereas $\mathbold{M_2(x)}$ represents the total momentum of the target signal plus uncertainties. Eq. \ref{eq5} provides a quantitative measure of the worth of an experiment, which guides the optimization of experimental design.

\subsection{Optimizing the Experimental Design}

The objective of experimental design optimization is to find the condition that would maximize $I(\mathbold{x})$ within the feasible domain of experimental conditions (usually constrained by the availability of diagnostics or physical limitations of the shock tube facility), i.e. $\mathbold{x}_{opt} = \argmax I(\mathbold{x})$. Generally, this problem does not have an analytical solution and needs to be solved numerically. A brute-force search method is used in the current work to find $\mathbold{x}_{opt}$.

\subsection{Analyzing Experimental Results}

As mentioned above, the art of rate constant analysis and uncertainty quantification has been explored extensively in various UQ studies and is not the primary focus of the current work . Nonetheless, a brief mathematical description of the current approach is provided here. Based on the Gaussian PDF assumption, the joint conditional probability distribution of $(\mathbold{\kappa},\mathbold{\xi})$ for a given data set $\mathbold{d}$ can be described as $prob(\mathbold{\kappa},\mathbold{\xi}|\mathbold{d}) \propto exp(-\Phi/2)$, where

\begin{equation}
\label{eq6}
\Phi = \frac{1}{\eta N} \sum_{i=1}^{N} \Big[ \frac{d(t_i) - f_{M}(t_i,\mathbold{x};\mathbold{\kappa},\mathbold{\xi})}{\sigma_{y}}\Big]^2 + \mathbold{\xi}^T\mathbold{\Sigma}^{-1}_{\mathbold{\xi}}\mathbold{\xi} + \mathbold{\kappa}^T\mathbold{\Sigma}^{-1}_{\mathbold{\kappa}}\mathbold{\kappa}
\end{equation}

Hence, determining the rate constants is mathematically equivalent to minimizing $\Phi(\mathbold{\kappa},\mathbold{\xi})$. This relation can be further generalized to a set of experiments investigating the same kinetic system with the same observable and data type, but performed at different conditions (e.g. repeated shock wave experiments at different temperatures and pressures):

The uncertainty in the inferred rate constants is described by the posterior PDF of $\mathbold{K}$ calculated by marginalizing $prob(\mathbold{K},\mathbold{\Xi}|\mathbold{D})$ over $\mathbold{\Xi}$. In well-designed experiments, the posterior PDF of $(\mathbold{\kappa},\mathbold{\xi})$ is usually sharply peaked, and a local Taylor series approximation can be applied near the minimal point, i.e. $(\mathbold{\kappa^*},\mathbold{\xi^*}) = \argmin(\Phi)$. Under this local approximation, it can be shown that $\mathbold{\kappa} \sim \mathcal{N}(\mathbold{\kappa^*},\mathbold{\Sigma_{\kappa|d}})$, where the posterior co-variance matrix $\mathbold{\Sigma_{\kappa|d}}$ can be determined from the Hessian matrix of $\Phi$ with respect to $(\mathbold{\kappa},\mathbold{\xi})$, i.e. $\mathbold{\Sigma_{\kappa|d}}$ equals the the upper left block of $2 (\nabla\nabla \Phi)^{-1}$ evaluated at $(\mathbold{\kappa^*},\mathbold{\xi^*})$. 

In the current work, minimization of $\Phi$ is solved numerically using the sequential quadratic programming (SQP) algorithm \cite{powell1983variable}, and a $\pm 3\sigma$ bounding box was imposed on $(\mathbold{\kappa},\mathbold{\xi})$ to improve the stability of the numerical solution. The numerical criterion for convergence was set to be $|\Delta\Phi| < 1 \times 10^{-2}$ between two consecutive iteration steps. Further discussion and optimization of the numerical recipe are reserved for future work.

\section{Example Study: Methanol Pyrolysis}
\begin{figure}
    \centering
    \includegraphics[width=0.5\linewidth]{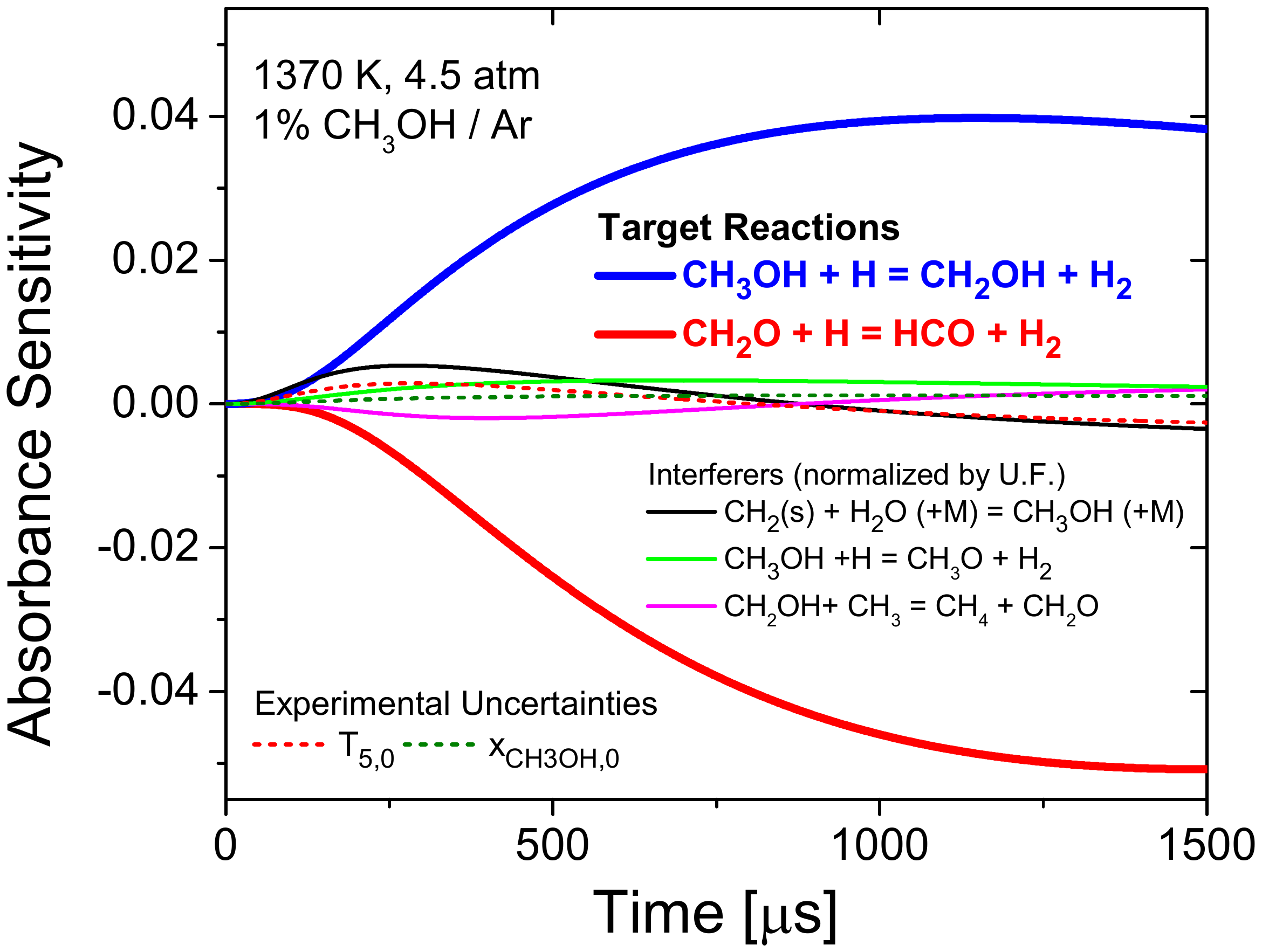}
    \caption{CH$_2$O absorbance sensitivity at the optimal experimental condition. Note that the sensitivity is dominated by the target rate constants $k_1$ and $k_2$.}
    \label{fig:F3}
\end{figure}

To illustrate the current method of experimental design and rate constant determination, a demonstrative shock tube study of the methanol pyrolysis system was conducted using diode laser absorption measurements of formaldehyde (CH$_2$O) time-histories. In this reaction system, the CH$_2$O concentration time-history is dominated by two reactions, namely CH$_2$O + H = HCO + H$_2$ (R1) and CH$_3$OH + H = CH$_2$OH + H$_2$ (R2) (see sensitivity analysis shown in Fig. \ref{fig:F3}). Using the information-based experimental design method, the current work identified the optimal experimental conditions for the simultaneous measurement of k$_1$ and k$_2$, then performed experiments near this optimal condition, and finally determined the two rate constants with unprecedented accuracy.

The experiments were conducted behind reflected shock waves in Stanford's 14.13-cm diameter Kinetic Shock Tube (KST) facility. Details about this facility have been documented in previous studies \cite{wang2015high,wang2017shock} and are not repeated here. Absorption measurements were performed using a recently developed 5.6 $\mu m$-laser diagnostic that has been optimized for CH$_2$O detection in combustion gases. Further details about this diagnostic can be found in \cite{ding2019formadelhyde}. To ensure data quality and minimize the influence of non-ideal effects in shock tubes, the current work only utilized absorbance signal in the first 1.5 ms of the total test time, although effective test times of 3 ms or longer were routinely achieved.

A compact reaction mechanism, namely the Foundational Fuel Chemistry Model (FFCM-1) \cite{FFCM}, was used as the trial kinetics model. The prior PDF of K (mean and co-variance matrix) was also obtained from FFCM-1. Based on the sensitivity analysis, the top 20 reactions were selected as the active reactions whose rate constants were adjustable, whereas the remaining reaction rates remained unperturbed. The random noise in the absorption signal, $\sigma_y$, was assumed to be $1.0 \times 10^{-3}$, which is consistent with the minimum detectable absorbance (MDA) obtained in most previous Stanford studies. The systematic experimental uncertainties ($1\sigma$) were estimated as follows: $\pm 0.5\%$ and $\pm 2\%$ in the reflected shock temperature and pressure, respectively; $\pm 2 \mu s$ in time zero; $\pm 0.001$ in the absorption base line; $\pm 2\%$ in the initial concentration of methanol; $\pm 2.5\%$ in the absorption cross-sections of CH$_2$O. The effect of residual impurities in the shock tube and mixing assembly was also considered and modeled as additional H radicals in the initial mixture composition -- an estimation method established in previous shock tube kinetics studies \cite{hong2011new,wang2017shock,mulvihill2019concerning}. The amount of H-equivalent impurities in Stanford's shock tube facilities and its temperature dependence has been studied previously by Urzay et al. \cite{urzay2014uncertainty} and modeled with a lognormal distribution, i.e. $ln(\chi_{H,0}) \sim \mathcal{N}(ln A_H - T_H/T, \sigma^2_H)$, where $A_H$ = 682.93 ppm, $T_H$ = 12,916 K and $\sigma_H$ = 1.86. The same value was assumed in the current study. The probability distribution of different sources of uncertainty were assumed to be mutually independent (i.e. $\mathbold{\Sigma_\xi}$ is a diagonal matrix). With $\mathbold{F_M}(\mathbold{x};\mathbold{\kappa},\mathbold{\xi})$, $\sigma_y$, $\mathbold{\Sigma_\kappa}$ and $\mathbold{\Sigma_\xi}$ fully defined, the experimental design was then optimized according to $I(\mathbold{x})$ and the rate constant measurements were performed subsequently, as discussed in the next two sections.

\begin{figure}
    \centering
    \includegraphics[width=0.5\linewidth]{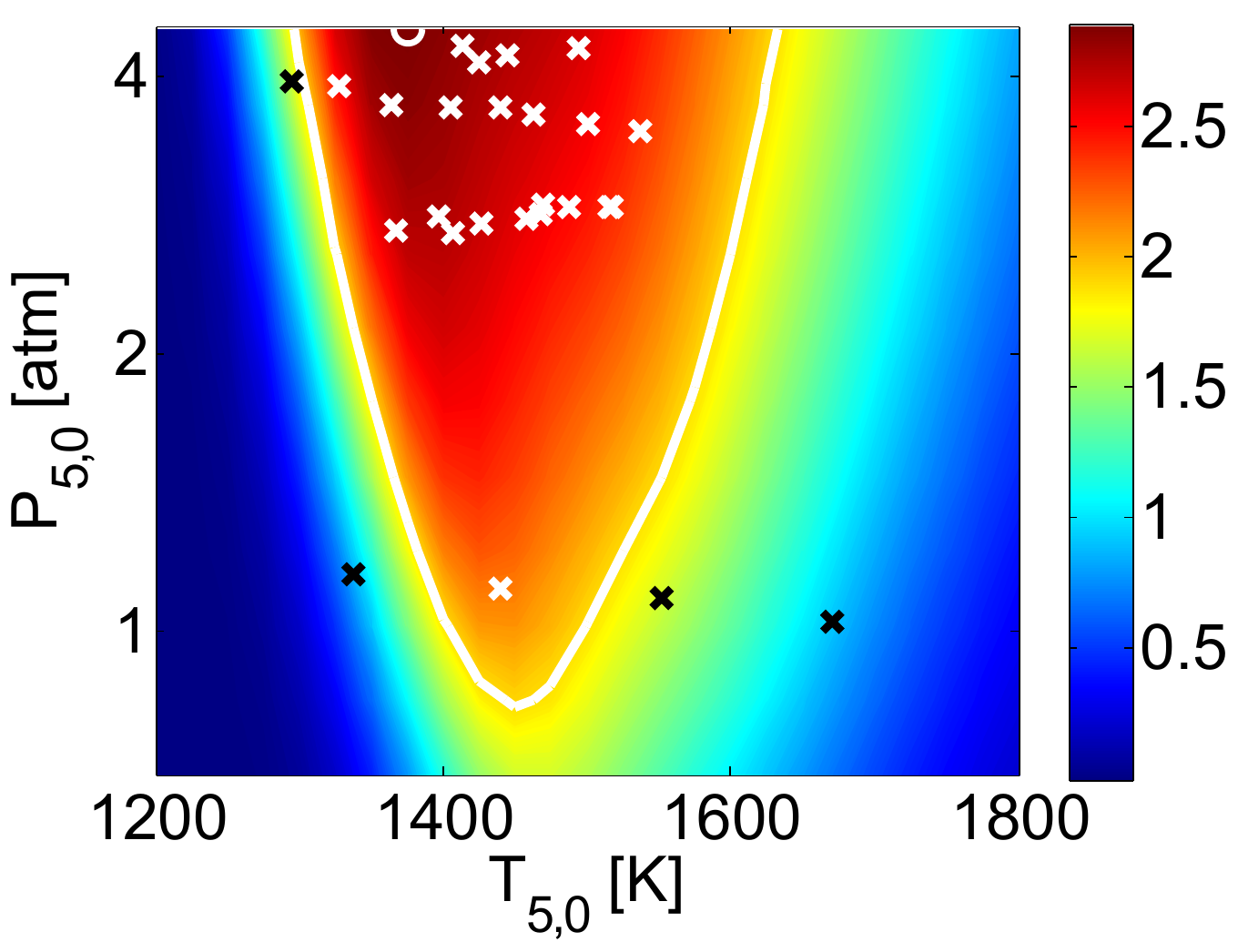}
    \caption{Expected information gain from CH$_2$O($t$) measurements at various reflected shock temperatures ($T_{5,0}$) and pressures ($P_{5,0}$). The optimal experimental condition is marked by the "o" symbol, whereas the actual conditions explored in the current experiments are marked by "x". The white line defines the boundary where the expected information gain is within 1 bit from the optimal value.}
    \label{fig:F2}
\end{figure}

The entire working range of the CH$_2$O diagnostic, i.e. 870 K $\le T_{5,0} \le$ 1800 K and 0.7 atm $\le P_{5,0} \le$ 4.5 atm, was investigated in the experimental design step. Within this feasible range of experimental conditions, $I(\mathbold{x})$ exhibits a positive dependence on the initial concentration of CH$_3$OH. However, in order to avoid condensation in the shock tube, the maximum concentration of CH$_3$OH was limited to 1 \%. For the 1\% CH$_3$OH / Ar mixture, Fig. \ref{fig:F2} shows the expected information gain is plotted as a function of $T_{5,0}$ and $P_{5,0}$. The optimal experimental condition was found to be $T_{5,0}$ = 1370 K and $P_{5,0}$ = 4.5 atm. A differential sensitivity analysis for the CH$_2$O absorbance signal (with respect to both the target reactions and systematic uncertainties) was also performed at this condition, as illustrated in Fig. \ref{fig:F3}. The majority of current experiments were conducted in the neighborhood of this condition where the expected information gain was within 1 bit from the optimal value. For comparison, a few experiments were also conducted far away from the optimal condition, yet the resulting data were not included in rate constant determination. All of the time-history data obtained in the current study have been made available in the Supplemental Materials.

\begin{figure}
    \centering
    \includegraphics[width=0.5\linewidth]{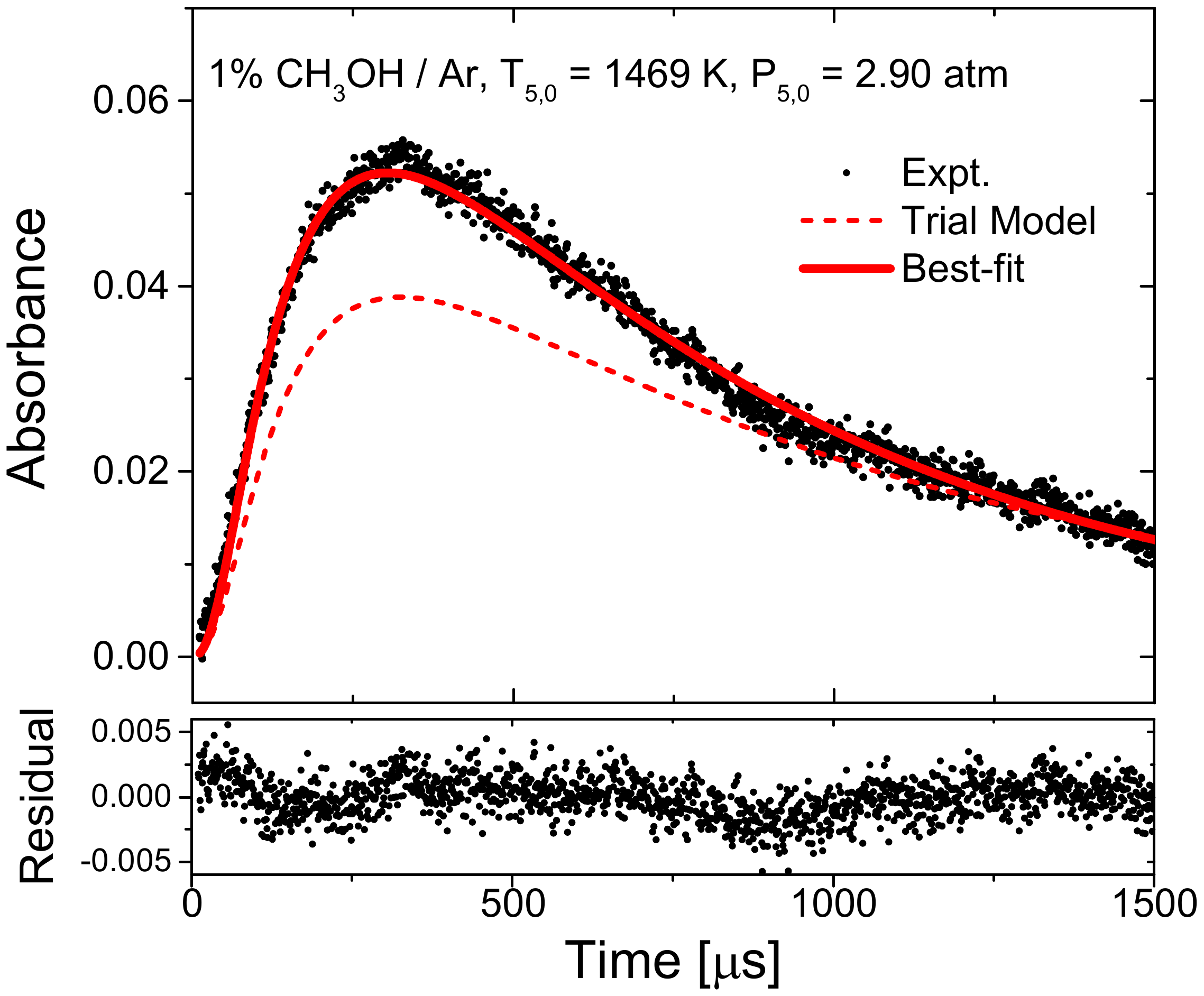}
    \caption{Representative CH$_2$O absorbance time-history data.}
    \label{fig:F4}
\end{figure}

\begin{figure}
    \centering
    \includegraphics[width=0.4\linewidth]{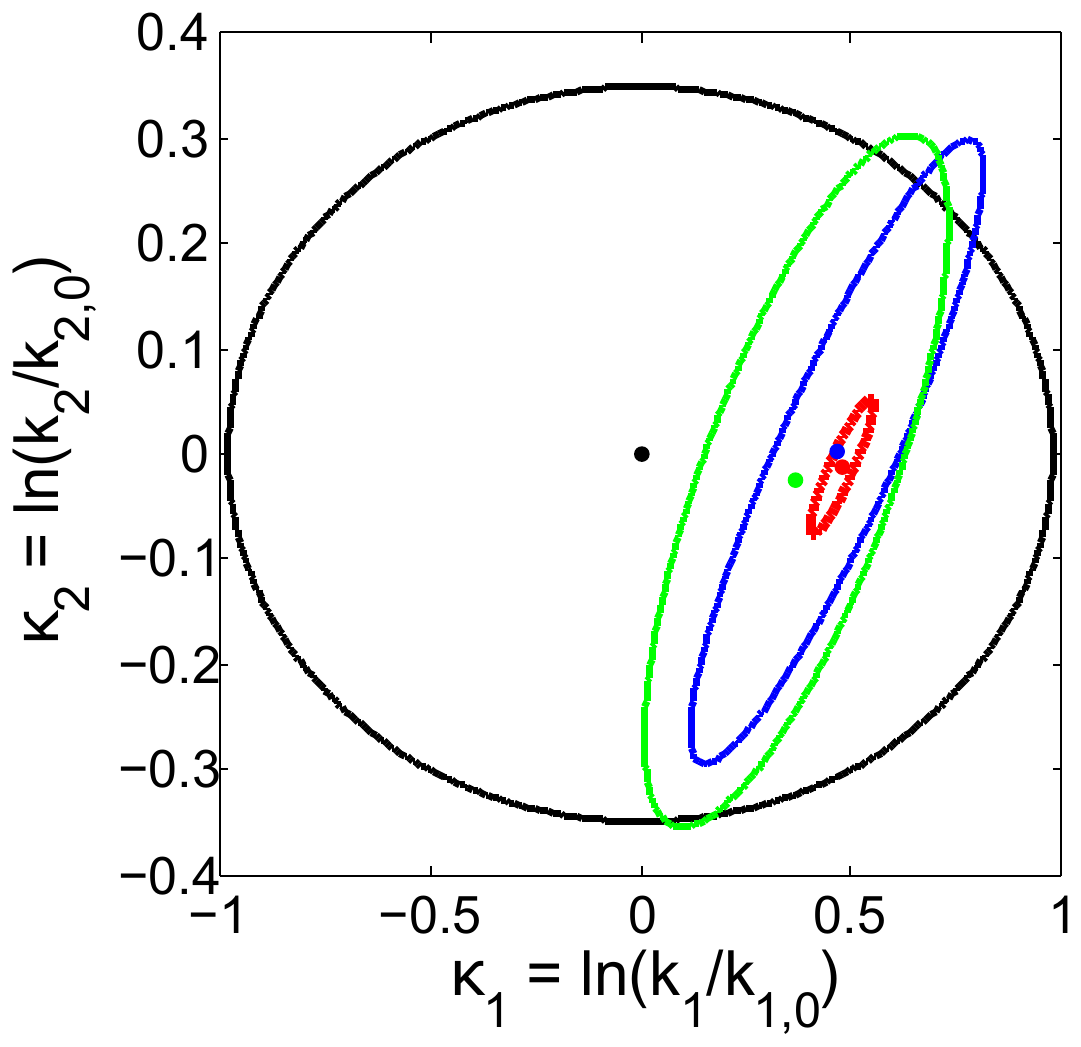}
    \caption{$2\sigma$ contours of $(\kappa_1,\kappa_2)$ probability densities. Black: prior distribution; blue: posterior distribution after a high-information experiment; green: posterior distribution after a low-information experiment; red: posterior distribution after a series of near-optimal experiments. The mean value of each distribution is indicated by the dot symbol.}
    \label{fig:F5}
\end{figure}

\begin{figure}
    \centering
    \includegraphics[width=0.5\linewidth]{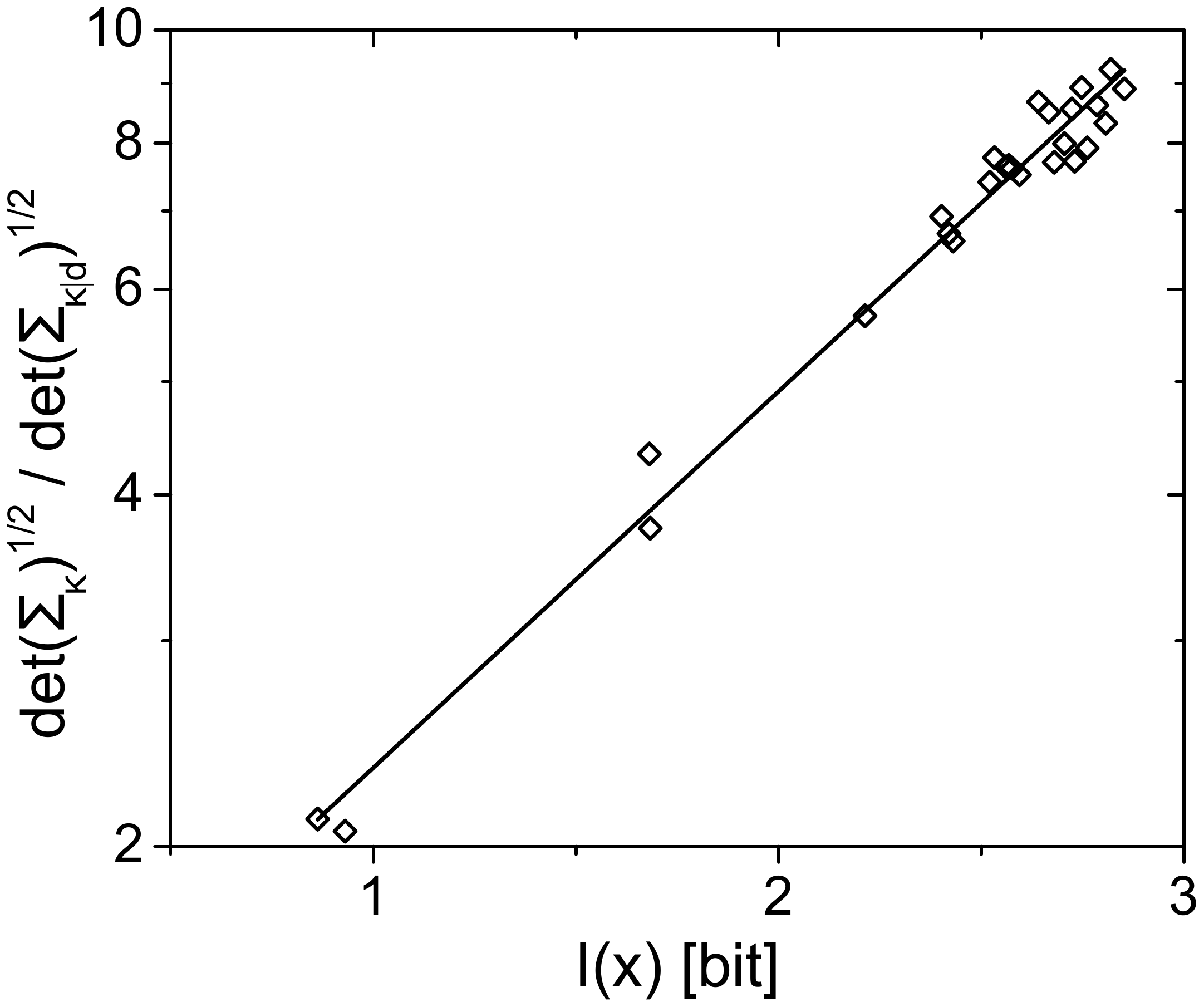}
    \caption{Actual vs. expected information gain from individual experiments.}
    \label{fig:F6}
\end{figure}

A representative CH$_2$O absorbance time-history measured in the current study is shown in Fig. \ref{fig:F4}. Also shown in the figure is the best-fit result calculated based on Eqn. \ref{eq4}. This particular experiment yields a set of normalized rate constant value as $(\kappa_1,\kappa_2)$ = (0.467, 0.002). The posterior PDF of $(\kappa_1,\kappa_2)$ (estimated based on local Hessian matrix of $\Phi$ in Eqn. \ref{eq6}) is shown in Fig. \ref{fig:F5}. Also shown in the figure is the result of a non-optimized experiment. Evidently, the $2\sigma$ contour of the non-optimized case is wider, meaning that it is less efficient in determining the target rate constants values. The overall uncertainty in the inferred rate constants can be measured by the area enclosed by the $2\sigma$ contour, which is proportional to $det(\mathbold{\Sigma_{\kappa|d}})^{1/2}$, and the actual information gain can be measured by the area ratio of the prior and posterior $2\sigma$ contours. Fig. \ref{fig:F6} compares the actual information gain from individual measurements, $det(\mathbold{\Sigma_{\kappa|d}})^{1/2} / det(\mathbold{\Sigma_{\kappa}})^{1/2}$, with the expected information gain $I(\mathbold{x})$ calculated before the experiments. A consistent trend is observed that for every 1 bit of increase in $I(\mathbold{x})$, the actual information gain approximately increases by a factor of 2, which confirms that the current definition of $I(\mathbold{x})$ is a good metric for predicting the benefit of an experiment.

The inferred rate constants $\kappa_1$ and $\kappa_2$ are seen to be correlated, as indicated by the ellipticity of the $2\sigma$ contour of their posterior PDF, because the sensitivities of CH$_2$O(t) to R1 and R2 ($\mathbold{S}_{\kappa1}$ and $\mathbold{S}_{\kappa2}$, respectively) are both strong. However, as long as $\mathbold{S}_{\kappa1}$ and $\mathbold{S}_{\kappa2}$ are linearly independent, the overall uncertainty in ($\kappa_1$, $\kappa_2$) could be further reduced through repeated measurements. A series of 22 experiments at near-optimal conditions were conducted in the current study (over the temperature range of 1287 - 1537 K), which led to a tight ultimate posterior PDF as shown in Fig. \ref{fig:F5}. The most probable values of $\kappa_1$ and $\kappa_2$ were found to be 0.478 and -0.012, respectively. Their uncertainties were also estimated by projecting the $2\sigma$ contour to individual axes. The resulting rate constant expressions are: $k_1 = 2.50 \times 10^6 (T/K)^{2.35} exp(-2975 K /T) \, cm^3mol^{-1}s^{-1} \pm 11.4\%$ and $k_2 = 7.06 \times 10^7 (T/K)^{1.9} exp(-1380 K/T) \, cm^3mol^{-1}s^{-1} \pm 9.7 \%$.

\begin{figure}
    \centering
    \includegraphics[width=0.5\linewidth]{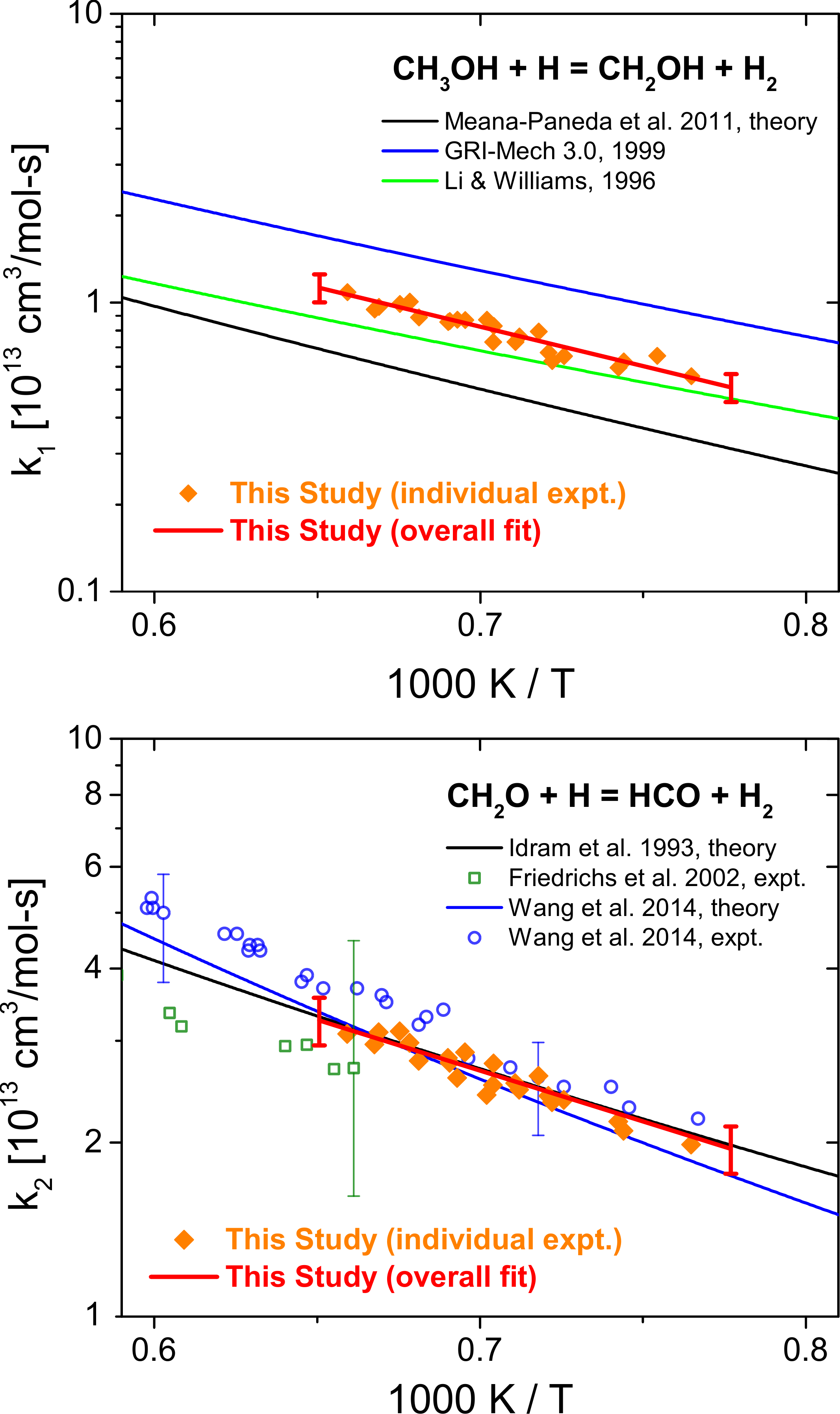}
    \caption{Arrhenius plots for the rate constants $k_1$ and $k_2$. Error bars represent the $2\sigma$ uncertainty limits of the experimental data.}
    \label{fig:F7}
\end{figure}

The current rate constant results are compared with previous theoretical and experimental studies in Fig. \ref{fig:F7}. For the reaction of CH$_3$OH + H = CH$_2$OH + H$_2$ (R1), results from the previous experimental study by Li and Williams in two-stage methanol flames \cite{li1996experimental} agrees well with the current rate constant data. The high-level direct-dynamics variational transition-state-theory calculation by Meana-Pañeda et al. \cite{meana2011high} (which is the value adopted in the trial model, FFCM-1) underpredicts the current results by 35\%, whereas the GRI-Mech 3.0 \cite{smith2011gri} expression overpredicts the current results by approximately 60\%. The rate constan of CH$_2$O + H = HCO + H$_2$ (R2) has been relatively well-studied both theoretically and experimentally by various researchers \cite{irdam1993formaldehyde, friedrichs2002direct,wang2014reaction}. The current results are seen in excellent agreement with previous experimental data (within their respective error bars)
and theories. The current work has significantly reduced the uncertainty in $k_2$ (by approximately a factor of 2 compared to the best previous result).

\section{Conclusions, Discussion and Outlook}
A new methodology and theoretical framework was established to aid the experimental design of shock tube / laser absorption studies of fundamental kinetic rate constants. Initial demonstration of this method in methanol pyrolysis system has proved successful and improved the rate constant determination of two key reactions, namely CH$_3$OH + H = CH$_2$OH + H$_2$ and CH$_2$O + H = HCO + H$_2$.

Although the current method was illustrated only with absorption time-history data, extension to other types of shock tube data should be straightforward under different definition of $\mathbold{d}$ and $\mathbold{F_M}$. The current method also accepts combinations of multiple data types (i.e. ignition delay time plus species time-history, or absorption time-histories of multiple species), via concatenation of these data into a long vector $\mathbold{d}$.

The objective of the experimental design can also be extended beyond measurements of a few specific rate constants. For example, experiments aimed at global uncertainty minimization of a reaction mechanism can be optimized based on the collective information gain in all reactions. The current method may also help designing experiment to improve prediction of physical quantities $\mathbold{Q}$ such as flame speed, lean blow-out limits and ignition delay times, which are functions of reaction rate constants. In such cases, the information gain should be calculated regarding the probability density of $\mathbold{Q = Q(K)}$.

It should be noted that current work did not consider model-form error such as missing reaction paths or species. Care is advised when applying the current experimental design method to a kinetic model that is incomplete (missing key reaction paths or species), not properly constrained or largely mispredicting the actual physics, as the local linearization assumption may no longer be valid. For best use of the current method, it is important to start with a good trial model that is close to ground-truth, since the mathematical expectation of information gain, as well as the inferred rate constants, depends on the prior knowledge of $\mathbold{K}$. Nonetheless, the influence of prior PDF diminishes as the number of experiments increases, and an iterative process of experimental design, execution and model update should, in principle, converge to a result that is independent to the initial model. Advanced topics such as asymptotic behavior, detection of model-form error, model comparison and selection, rejection of spurious data and optimization of numerical algorithm, are reserved for future studies.

\section*{Acknowledgments}
This work is supported by the U.S. Army Research Office under Grant No. W911NF-17-1-0420, with Dr. Ralph Anthenien as the contract monitor, and also by the Air Force Office of Scientific Research through AFOSR Grant No. FA9550-16-1-0195, with Dr. Chiping Li as the contract monitor.

\bibliography{sample.bbl}
\bibliographystyle{elsarticle-num-PROCI.bst}

\end{document}